\documentclass{article}

\usepackage{amsmath}
\usepackage{graphicx}
\usepackage{enumerate}
\usepackage{natbib}
\usepackage{amsmath, calc}
\usepackage{booktabs}
\usepackage{multirow}
\usepackage{rotating, tabularx}
\usepackage{url} 
\usepackage{subfigure}
\usepackage{float}
\usepackage{authblk}
\usepackage[margin=1in]{geometry}
\usepackage{changes} 

\title{A Bayesian Joint Modelling for Misclassified Interval-censoring and Competing Risks}

\author[1,2]{Zhenwei Yang}
\author[1,2]{Dimitris Rizopoulos}
\author[3]{Eveline A.M. Heijnsdijk}
\author[4]{Lisa F. Newcomb}
\author[1,2]{Nicole S. Erler}
\affil[1]{Department of Biostatistics, Erasmus Medical Center Rotterdam}
\affil[2]{Department of Epidemiology, Erasmus Medical Center Rotterdam}
\affil[3]{Department of Public Health, Erasmus Medical Center Rotterdam}
\affil[4]{Fred Hutchinson Cancer Center, Cancer Prevention Program, Public Health Sciences, Seattle, Washington}

\begin{document}

\maketitle

\begin{abstract}
In active surveillance of prostate cancer, cancer progression is interval-censored and the examination to detect progression is subject to misclassification, usually false negatives. Meanwhile, patients may initiate early treatment before progression detection, constituting a competing risk. We developed the Misclassification-Corrected Interval-censored Cause-specific Joint Model (MCICJM) to estimate the association between longitudinal biomarkers and cancer progression in this setting. The sensitivity of the examination is considered in the likelihood of this model via a parameter that may be set to a specific value if the sensitivity is known, or for which a prior distribution can be specified if the sensitivity is unknown. Our simulation results show that misspecification of the sensitivity parameter or ignoring it entirely impacts the model parameters, especially the parameter uncertainty and the baseline hazards. Moreover, specification of a prior distribution for the sensitivity parameter may reduce the risk of misspecification in settings where the exact sensitivity is unknown, but may cause identifiability issues. Thus, imposing restrictions on the baseline hazards is recommended. A trade-off between modelling with a sensitivity constant at the risk of misspecification and a sensitivity prior at the cost of flexibility needs to be decided.
\end{abstract}



\maketitle

\section{Introduction}
\label{sec:intro}
In cancer research, the relation between repeated biomarker measurements and the risk of cancer-related events, such as the progression of the cancer to a stage that requires an intervention,  is usually of great interest.\citep{Pepe2001, Hemelrijck2011, Putt2015} The repeatedly measured biomarkers typically have short-term fluctuations and are intrinsically associated with patients' status, in other words, endogenous.\citep{Dimitris2011} Joint models for longitudinal and time-to-event outcomes have been developed to overcome these challenges.\citep{Dimitris2012} Cancer-related events have the additional difficulty that they are usually silent and detected by periodic, and often imperfect examinations. In the motivating Canary Prostate Active Surveillance Study (PASS), the primary event of interest, cancer progression, is never exactly observed but is usually detected by (ir)regular prostate biopsies, and thus interval-censored.\citep{EPSTEIN20121019} Moreover, biopsies, whose sensitivity of detecting cancer progression is estimated to be as low as 50-90\%\citep{Inoue2014, Tiago2017, Lange2020}, are subject to false-negative results, leading to misclassification. This means that cancer progression may remain undetected even though multiple biopsies are performed over time. The situation can be further complicated if competing risks exist. For instance, around 10\% of the patients in the Canary PASS cohort started treatment before cancer progression was detected.\citep{Newcomb2016} Since the start of treatment means the end of active surveillance, the endpoint of interest, cancer progression while on active surveillance, can no longer be observed and early initiation of treatment, thus the treatment initiation constitutes a competing risk. In such cases, the standard joint models are no longer valid.

Previous studies have only addressed some of the above-mentioned challenges. \citet{Lee2020} proposed a prediction model, composed of a recurrent neural network and cause-specific neural networks to handle the repeatedly measured biomarkers and competing events. \citet{Tomer2022} extended the standard joint models to handle one interval-censored event, but could not handle competing risks and the imperfect sensitivity. We previously extended the joint model further to handle interval-censored events along with exactly-observed competing events.\citep{Yang2023} None of these approaches, however, considered the possible misclassification of the time-to-event outcome. \citet{Magda2021} elaborated on the methodology of incorporating specificity and sensitivity in survival models, however, only considered time-constant exogenous covariates to handle a single potentially misclassified interval-censored event. 

In the present paper, we address the issue of outcome misclassification due to false-negative biopsies, which has not been considered in the existing literature. We introduce an extra sensitivity parameter to our previously proposed model for right- and interval-censored competing risk joint models, and formulate this Misclassification-Corrected Interval-censored Cause-specific Joint Model (MCICJM, the corresponding code can be found on \url{https://github.com/ZhenweiYang96/Misclassification-MCICJM}). In this new model, the sensitivity parameter can be set to a pre-specified, fixed constant. Alternatively, since the exact value of the sensitivity is typically unknown in clinical practice, a prior distribution can be specified for the sensitivity parameter. This extension is not only relevant for obtaining unbiased estimates of the hazard ratios from the joint model, but is also relevant in the context of (dynamic) prediction using this model, as the modelling of the biopsy sensitivity influences the estimation of the baseline hazard.

The remainder of this paper is structured as follows. In Section~\ref{sec:data}, we describe the motivating Canary PASS data, the basic structure of the joint model used to analyse this data, and the misclassification issue in this application. Section~\ref{sec:method} defines the proposed MCICJM. The results from the analysis of the Canary PASS data using the MCICJM considering three fixed biopsy sensitivities are presented in Section~\ref{sec:analysis}. Section~\ref{sec:simulation} presents a simulation study for the performance of the MCICJM under different scenarios with different (misspecified) fixed sensitivity parameters or priors for the biopsy sensitivity. Section~\ref{sec:discussion} concludes this paper with a discussion.

\section{Real World Case}
\label{sec:data}

\subsection{The Canary PASS Data}

The Canary PASS is a multicentre active surveillance (AS) study launched in 2008 by the Canary Foundation and the National Cancer Institute.\citep{Newcomb2016} Included are previously untreated patients who were diagnosed with low-risk prostate cancer (i.e., a Gleason score of six) and for whom a confirmatory biopsy showed a Gleason score of six or no cancer.\citep{Cooperberg2020} Recruited patients are required to undergo prostate-specific antigen (PSA) tests every three months, clinical visits every six months, and biopsies at 12 and 24 months, and biennially afterwards. The present study includes a subset of 833 patients recruited until 2017 who had at least one PSA measurement during AS. The primary outcome, cancer progression, is defined as progression to a Gleason score of seven or higher. Patients observed with this endpoint leave AS to receive active treatment (e.g., radiotherapy or prostatectomy). Notably, 87 patients left AS for early treatment before cancer progression was detected, constituting a competing event. The data is summarized in Table S1.

Previous literature indicated that the PSA value along with its dynamic change (e.g., yearly change) are essential predictors of cancer progression.\citep{Nelson2021, Cary2013} 
Likewise, higher PSA values or rapid PSA growth may alert the patient and/or doctor and ultimately lead to the initiation of early treatment. Thus, in this study, we explore the impact of PSA on cancer progression and early treatment as a competing event. The longitudinal outcome, PSA, is visualized for 20 randomly selected patients in Figure S2.

\subsection{Joint Modelling}

In the context of repeatedly measured biomarkers and the risks of two competing events, the framework of the joint models (JM) for longitudinal and time-to-event data is an established method, and is gaining popularity in applied research. Moreover, the JM is suitable for predicting cancer progression based on the longitudinal history of PSA values. The JM consists of two parts, the longitudinal sub-model capturing the underlying trajectory of PSA measurements and the time-to-event sub-model modelling the hazards of (underlying) cancer progression and early treatment initiation. The two parts are linked via shared random effects, $\boldsymbol{u}_i$. Specifically for the PASS data, the longitudinal part is, in essence, a generalized mixed model, in which the outcome is the logarithm of the PSA values which, conditional on covariates and the random effects, is assumed to have nonlinear evolutions over time are captured by a natural cubic spline (three degrees of freedom). The time-to-event sub-model is a time-dependent cause-specific proportional hazard model incorporating different functional forms of the time-varying PSA measurements and additional covariates. By including functions of the estimated trajectory of the longitudinal biomarker(s) the short-term fluctuations of the measured marker values, which are typically not of interest and may obscure the underlying association, are filtered out. The model can be written as follows:
\begin{align*}
\begin{cases}
    &\textbf{Longitudinal sub-model:} \\
    &\log_2\{\mbox{PSA}_i(t) + 1\} = m_i(t) + \epsilon_i(t) = \beta_0 + u_{0i} + \displaystyle{\sum}^3_{p=1}(\beta_p + u_{pi})\mathcal{C}^{(p)}_i(t)  + \beta_4(\mbox{Age}_i - 62) + \epsilon_i(t), \quad \boldsymbol{u}_i \sim \mathcal{N}(\mathbf{0}, \boldsymbol{\Omega}), \\
    &\textbf{Time-to-event sub-model:} \\
    &h_i^{(k)}\left\{t \mid \boldsymbol{\mathcal{M}}_{i}(t), \mbox{PSAD}_i\right\} = h_0^{(k)}(t)\exp\left[\gamma_k^\top\mbox{PSAD}_i + f_{k}\{\boldsymbol{\mathcal{M}}_{i}(t),  \boldsymbol{\alpha}_{k}\}\right],
\end{cases}
\end{align*}
where $m_i(t)$ is estimated PSA value of patient $i$ at time $t$;  $\mathcal{C}^{(1)}(\cdot), \dots, \mathcal{C}^{(3)}(\cdot)$ are the set of basis functions for the natural cubic splines with three degrees of freedom  (without an intercept);; $\mbox{Age}_i$ refers to the patient's age at the start of active surveillance; and $\boldsymbol{\beta} = (\beta_0, \dots, \beta_4)^\top$ is a vector of corresponding regression coefficients. Following \citet{Tomer2018}, the error terms $\epsilon_i(t)$ are assumed to follow a Student's t-distribution (with three degrees of freedom). The random effects, $\boldsymbol{u}_i = (u_{0i}, u_{1i}, u_{2i}, u_{3i})^\top$, are assumed to be normally distributed with a mean of zero and an unstructured variance-covariance matrix $\boldsymbol{\Omega}$. The hazard of the $i$-th patient for event $k$ ($k \in \boldsymbol{\mathcal{K}}=\{\textsc{prg}, \textsc{trt}\}$, where $\textsc{prg}$ stands for cancer progression and $\textsc{trt}$ for early treatment) at time $t$ is denoted as $h_i^{(k)}(t)$. The vector $\boldsymbol{\mathcal{M}}_{i}(t) = \{m_{i}(s); 0 \leq s < t\}$, contains the history of the estimated trajectory of PSA until $t$, and $\mbox{PSAD}_i$ is the baseline PSA density (calculated as PSA divided by prostate volume at baseline), with corresponding regression coefficient $\gamma_k$. We denote $f_{k}(\cdot)$ as the functional form of the longitudinal outcome (i.e., PSA) on the hazard of event $k$, and $\boldsymbol{\alpha}_{k}$ is the corresponding parameter vector. The event-specific baseline hazard, $h_0^{(k)}(t)$, is typically modelled using P-splines.

Consistent with our earilier work\citep{Yang2023}, we consider two functional forms of PSA, the expected value at time $t$ (``$\log_2(\text{PSA} + 1)$ value") and the expected change over the year before time $t$ (``$\log_2(\text{PSA} + 1)$ yearly change"), i.e.,
\begin{align*}
    f_{k}\left\{\boldsymbol{\mathcal{M}}_{i}(t), \boldsymbol{\alpha}_{k}\right\} = \alpha_{1k} m_{i}(t) + \alpha_{2k}\{m_{i}(t) - m_{i}(t - 1)\}.
\end{align*}
Other options of functional forms may include the instantaneous slope change of PSA, which is $m'_i(t)$, the first derivate of $m_i(t)$. However, this slope change is less clinically interpretable than the ``yearly change".

\subsection{Challenges of modelling cancer progression}

Cancer progression is detected by periodic biopsies and cannot be observed directly. To avoid bias, cancer progression must be considered as interval-censored by the times of the biopsy at which progression was detected and the previous, negative biopsy.  \citet{Tomer2022} developed a joint model to deal with this challenge. In practice, however, the situation is further complicated by the fact that the biopsies used to determine cancer progression are error-prone. The small tissue sample obtained by the biopsy may not contain cells from the most affected portion of the prostate\citep{Serefoglu2013}, leading to an underestimation of the Gleason score and thereby causing a false-negative result. Consequently, the true progression time could be earlier than the interval at which progression was detected. Treating cancer progression as a standard interval-censored event would assume event times that are typically too late and ignore the additional uncertainty about when progression happened. To avoid the ensuing bias, there is an urgent need to extend the joint modelling framework with the capability to handle misclassified event statuses. One way of doing this is to explicitly include the biopsy sensitivity in the model likelihood via a pre-specified constant, $\rho \in [0,1]$. An additional issue to be taken into account in this context is the uncertainty brought by the biopsy sensitivity since the biopsy sensitivity, $\rho$, is typically unknown in the real world.

\section{Estimation Accounting for Imperfect Biopsies}
\label{sec:method}

\subsection{Notation}
\label{subsec:notation}
In this section, the notation is demonstrated in the context of the PASS data. We denote $T^*_i$ as the true time of the event that happens first among the two events, i.e., cancer progression and early treatment, for the $i$-th patient, $T^*_i = \min\{T^{\textsc{prg*}}_i, T^{\textsc{trt*}}_i\}$. The censoring time is denoted as $T^{\textsc{cen}}_i$. Patient $i$ in the data experienced $N_i + 1$ biopsies in total, the time of which is denoted by $\boldsymbol{t}^{(b)}_i = \{t^{(b)}_{0,i}, t^{(b)}_{1,i}, \dots, t^{(b)}_{N_i,i}\}$, where $t^{(b)}_{0,i}$ corresponds to the time of the biopsy that is always performed at the start of AS. We denote the event indicator of patient $i$, $\delta_i = \{0, 1, 2\}$, where censored patients are represented by $\delta_i =0$, patients who are detected with progression $\delta_i=1$, and those who started early treatment $\delta_i=2$. If cancer progression was detected in patient $i$ ($\delta_i=1$), his observed event time (i.e., the progression detection time) is the time of the last (i.e., the $(N_i+1)$-th) biopsy, $T^{\textsc{prg}}_i=t^{(b)}_{N_i,i}$, after which he leaves the AS program. The primary event, cancer progression ($k=\textsc{prg}$), is interval censored by the $N_i$ biopsy intervals. Since biopsies may give false negative results, progression may actually have occurred in any of the prior biopsy intervals but was missed. For a patient in whom progression was detected ($\delta_i=1$), we, therefore, have a vector of potential progression-free times, which is the vector of left boundaries of the intervals defined by the biopsy times $\{t^{(b)}_{0,i},  \dots, t^{(b)}_{N_i-1,i}\}$ (i.e., excluding the last biopsy). For such a patient, it is known that treatment was not started until the end of follow-up, i.e., the time of the last biopsy, $t^{(b)}_{N_i,i}$, constitutes the treatment-free survival time. Patients who initiated early treatment ($\delta_i=2$) or were censored ($\delta_i=0$), also have a vector of potential progression-free times, since their cancer may have progressed but was missed by all biopsies. For these patients, the vector of potential progression-free times $\{t^{(b)}_{0,i}, t^{(b)}_{1,i}, \dots, t^{(b)}_{N_i,i}\}$, thus, includes the last biopsy. Contrary to patients detected with progression, for these two groups of patients the possibility that their cancer did not progress until the last (negative) biopsy has to be considered as well. The treatment-free time for a patient who started early treatment is the treatment initiation time (which is typically recorded in the medical records and therefore known exactly), $T^{\textsc{trt}}_i$, whereas the treatment-free time for a patient who was censored is the censoring time $T^{\textsc{cen}}_i$. In summary, the vector of the observed times for patient $i$ is given by
\begin{align*}
\boldsymbol{T}_i = 
    \begin{cases}
    \left(t^{(b)}_{0,i},  \dots, t^{(b)}_{N_i,i}, T^{\textsc{cen}}_i\right), &\text{if }\delta_i=0, \\
    \left(t^{(b)}_{0,i},  \dots, t^{(b)}_{N_i-1,i}, T^{\textsc{prg}}_i=t^{(b)}_{N_i,i}\right), &\text{if }\delta_i=1, \\
    \left(t^{(b)}_{0,i},  \dots, t^{(b)}_{N_i,i}, T^{\textsc{trt}}_i\right), &\text{if }\delta_i=2.
    \end{cases}
\end{align*}

The typically used transrectal ultrasound guided prostatic biopsy has a reported specificity of 96\%.\citep{Streicher2019} Therefore, for simplicity, we assume that the biopsies in our application do not yield false-positive results. Incorporating this reported specificity in our model would most likely not result in very different results. Note that here we also assume the first biopsy ($t^{(b)}_{0,i}$) is never a false-negative. This assumption could be relaxed by including an offset. This offset depends on the biopsy sensitivity as well as the risk of progression before the start of AS. 

The vectors of observed event times, $\boldsymbol{T}_i$, together with $n_i$ longitudinal PSA measurements $\boldsymbol{y}_i =\{y_{i,1}, \dots, y_{i,n_i}\}$, and covariates $\boldsymbol{X}_i$ such as age at the start of AS, PSA density and biomarker measurement times, form the observed data, $\boldsymbol{\mathcal{D}}_n = \{\boldsymbol{T}_i, \delta_i, \boldsymbol{y}_{i}, 
\boldsymbol{X}_i; i = 1, \dots, n\}$ for all $n$ patients.

\subsection{Misclassification-corrected Interval-censored Cause-specific Joint Models (MCICJM) for Longitudinal and Time-to-event Outcomes}
\label{subsec:mcicjm}

To correctly estimate the association between the PSA trajectories and the risk of cancer progression, the added uncertainty about the time of progression, due to potentially false-negative biopsies has to be taken into account. For this purpose, we introduce the parameter $\rho$ that denotes the biopsy sensitivity, i.e., the probability that a biopsy is positive when the cancer has progressed. For now, we assume the value of $\rho$ to be fixed (across patients and biopsies) and known a priori. The likelihood for patient $i$ can then be written as:
\begin{align*}
    p\left\{\boldsymbol{y}_{i}, \boldsymbol{T}_i, \delta_i \mid \boldsymbol{u}_i, \rho, \boldsymbol{\theta}\right\} = \prod^{n_{i}}_{\ell=1}p(y_{i,\ell}\mid \boldsymbol{u}_{i}, \boldsymbol{\theta}) \times p\left\{\boldsymbol{T}_i, \delta_i \mid \boldsymbol{u}_i, \rho, \boldsymbol{\theta}\right\},
\end{align*}
where $\delta_i$ and $\boldsymbol{T}_i$ are the event indicator and observed event times, respectively, as defined in the previous subsection, $n_{i}$ is the number of repeated measurements of the longitudinal outcome for patient $i$, and $\boldsymbol{\theta}$ is the vector of all model parameters (including the baseline hazard regression coefficients, covariates coefficients, fixed effect, residual variance and other hyperparameters) other than the biopsy sensitivity $\rho$ and the random effects $\boldsymbol{u}_i$. It is assumed that, conditional on the random effects $\boldsymbol{u}_{i}$, the longitudinal (i.e., the first factor of the likelihood formula) and survival (i.e., the second factor of the likelihood formula) part are independent, and the repeated measurements of the longitudinal outcomes for the same patient are independent. The likelihood for the longitudinal part is:
\begin{align*}
\prod^{n_{i}}_{\ell=1}p(y_{i,\ell}\mid \boldsymbol{u}_{i}, \boldsymbol{\theta}) = \prod^{n_{i}}_{\ell=1}\frac{\Gamma(\frac{\kappa+1}{2})}{\sqrt{\kappa \pi \sigma^2}\Gamma(\frac{\kappa}{2})}\left(1+\frac{(y_{i,\ell} - \boldsymbol{w}_{i,\ell}^\top \boldsymbol{\beta} - \boldsymbol{z}_{i,\ell}^\top\boldsymbol{u}_i)^2}{\kappa \sigma^2}\right)^{-\frac{\kappa+1}{2}},
\end{align*}
where $\Gamma(\cdot)$ is the Gamma function; $\kappa$ is the degrees of freedom of the t distribution (in our case, $\kappa = 3$); $\sigma$ is the standard deviation of the residuals; $\boldsymbol{w}_{i,\ell}$ and $\boldsymbol{z}_{i,\ell}$ are the vectors of fixed effects (in our case including an intercept, the natural cubic splines of time and baseline age) and random effects (in our case including an intercept and the natural cubic splines of time) for the $\ell$-th repeated measurement of patient $i$.

The likelihood of the survival part differs depending on the observed event type for each individual. Each component describes how likely the observed data are under the scenarios (regarding the true time of cancer progression and false-negative biopsy results) that would lead to the observed data. Specifically, the likelihood for the survival part can be written as:
\begin{align*}
    p(\boldsymbol{T}_i, \delta_i \mid \boldsymbol{u}_i, \rho, \boldsymbol{\theta}) = \mathcal{F}_1\times \mathcal{F}_2  \times \mathcal{F}_3, 
\end{align*}
where 
\begin{align*}
\mathcal{F}_1 = \Bigg[\exp\left\{- H_i^{(\textsc{prg})}(t^{(b)}_{N_i,i}) - H^{(\textsc{trt})}_i(T^{\textsc{cen}}_i)\right\} + \sum^{N_i}_{j=1} \mathcal{A}_{ij} (1 - \rho)^{N_i-j+1} \exp\left\{-H^{(\textsc{trt})}_i(T^{\textsc{cen}}_i)\right\} \Bigg]^{I(\delta_i = 0)},
\end{align*}
\begin{align*}
\mathcal{F}_2 = \left[\sum^{N_i}_{j=1} \mathcal{A}_{ij} \rho(1-\rho)^{N_i - j} \exp\left\{-H^{(\textsc{trt})}_i(t^{(b)}_{N_i,i})\right\}\right]^{I(\delta_i = 1)}, 
\end{align*}
\begin{align*}
\mathcal{F}_3  = \Bigg[h_i^{(\textsc{trt})}(T^{\textsc{trt}}_i)\exp\left\{- H_i^{(\textsc{prg})}(t^{(b)}_{N_i,i}) - H^{(\textsc{trt})}_i(T^{\textsc{trt}}_i)\right\} + 
\sum^{N_i}_{j=1} \mathcal{A}_{ij} (1 - \rho)^{N_i-j+1} h^{(\textsc{trt})}_i(T^{\textsc{trt}}_i)\exp\left\{-H^{(\textsc{trt})}_i(T^{\textsc{trt}}_i)\right\} \Bigg]^{I(\delta_i=2)},
\end{align*}
with $\mathcal{A}_{ij} = \displaystyle{\int}^{t^{(b)}_{j,i}}_{t^{(b)}_{j-1,i}}h_i^{(\textsc{prg})}(\nu)\exp\left\{-H^{(\textsc{prg})}_i(\nu)\right\}d\nu$. $H^{(\textsc{prg})}_i(\cdot)$ and $H^{(\textsc{trt})}_i(\cdot)$ are the progression- and treatment-specific cumulative hazard and $I(\cdot)$ denotes the indicator function. The first factor, $\mathcal{F}_1$, models the likelihood that patient $i$ is either event-free or that progression occurred but was never detected. Specifically, $\mathcal{A}_{ij}$ describes the probability that patient $i$'s cancer did progress in the interval $\left[t^{(b)}_{j-1,i}, t^{(b)}_{j,i}\right]$, and $(1 - \rho)^{N_i-j+1}$ is the probability that all $N_i-j+1$ biopsies $t^{(b)}_{j,i}, \dots,t^{(b)}_{N_i,i}$ after this interval gave false-negative results. The second factor, $\mathcal{F}_2$, describes the probability that progression (detected at the last biopsy) actually occurred in this or any of the previous biopsy intervals, and that treatment was not initiated before the last biopsy. The last part, $\mathcal{F}_3$, analogous to the first part, models the probability that patient $i$ started treatment at $T^{\textsc{trt}}_i$ and is either progression-free until the last biopsy $t^{(b)}_{N_i,i}$ or that progression was not detected. The integrals in the above align* do not have a closed-form solution and were numerically approximated using the seven-point Gauss-Kronrod rule. The model was estimated under the Bayesian framework using Markov Chain Monte Carlo (MCMC) methods, Gibbs sampling combined with the Metropolis-Hastings algorithm, in JAGS.\citep{JAGS}

\subsection{Uncertainty about the Sensitivity Parameter}
\label{subsec:sensunc}

Up to now, we have assumed the sensitivity of the biopsies ($\rho$) to be known. However, in clinical practice it is not realistic to assume the value of the sensitivity (or specificity) of a diagnostic test to be known without uncertainty. The Bayesian framework is generally very well suited to incorporate uncertainty about model parameters by a specification of a (hyper-)prior distribution. In the absence of an additional (gold-standard) measurement of a patient's true cancer progression status, it is not possible to infer the likely values of the biopsy sensitivity from the data. The prior distribution for $\rho$, thus, has to be informative. As an extra free parameter that is not estimable, the sensitivity parameter $\rho$ can challenge the identifiability of the MCICJM.\citep{Hausman1998, Magda2021} Previous literature suggested two ways to improve model identifiability: reducing the dimensionality of the parameter vector of the model,\citep{Balasubramanian2003, Torelli2002} and using a very informative prior distribution for the undefined parameter.\citep{Magda2021} In our case, the ambiguity of the underlying progression time (which is earlier than the observed time) likely applies to all three types of patients (see Section \ref{subsec:mcicjm}). Intuitively, the model should bring the estimated event time forward, which can be achieved by increasing the progression-specific baseline hazard. The baseline hazard, thus, interplays with the biopsy sensitivity. 

Typically, we model the baseline hazard in a joint model using P-splines with a high number of degrees of freedom.\citep{Eilers1996}  In combination with the inestimable sensitivity parameter, however, this may burden the model identifiability. Therefore, we restrict the dimension of the parameter vector in the P-spline specification of the baseline hazard by reducing the degrees of freedom from 12 to 5 (thereby limiting the flexibility of the shape of the baseline hazard). To reflect the uncertainty about the biopsy sensitivity, for which values between 50\% and 90\% are considered plausible,\citep{Inoue2014, Tiago2017, Lange2020} we chose a uniform distribution on this range as the prior for $\rho$. Alternatively, a beta distribution could be considered.

\section{Real Data Analysis}
\label{sec:analysis}

We fitted three MCICJMs with fixed biopsy sensitivities of 60\% and 80\% and a uniform prior between 50\% and 90\%, and compared the parameter estimates with those from the previously developed ICJM from \citet{Yang2023} (i.e., equivalent to the MCICJM assuming 100\% biopsy sensitivity). The most relevant parameter estimates are presented in Table \ref{tab:modelresult}. The effect sizes (i.e., coefficient estimates) of PSA density and the width of credible intervals were similar across the three models. Doubling the PSA density resulted in a cancer progression-specific risk increase by a factor of 1.31, 1.32, 1.36 and 1.28, when assuming 100\%, 80\%, 60\%, and the uniform prior, respectively. In addition, the impact of $\log_2(\text{PSA + 1})$ on both cancer progression and early treatment was comparable across models. One unit increase in $\log_2(\text{PSA + 1})$ led to a hazard ratio of 1.17-1.23. In contrast, with lower fixed sensitivity, the associations between the PSA yearly change and cancer progression were larger. A one-unit increase in the $\log_2(\text{PSA + 1}) \text{ yearly change}$ contributed to a 5.57-, 5.8-, 7.15, and 4.36-fold increase in the risk of cancer progression, with 100\%, 80\% and 60\% sensitivity, and a uniform prior between 50\% and 90\%, respectively. Similarly, the credible intervals were wider in the models with lower fixed sensitivities. On the other hand, there was no clear pattern in the treatment-specific hazard ratios of PSA density, $\log_2(\text{PSA + 1})$ or $\log_2(\text{PSA + 1}) \text{ yearly change}$. However, the uniform prior led to the lowest estimated hazard ratios of $\log_2(\text{PSA + 1}) \text{ yearly change}$, 5.36 for progression and 7.21 for early treatment, along with the narrowest credible interval width, compared among the four models.

\begin{table}[H]
    \setlength\extrarowheight{3pt}
    \caption{Estimated parameters in the joint models for the Canary PASS data.}
    \label{tab:modelresult}
    \resizebox{\textwidth}{!}{%
        \begin{tabular}{lcccccccc}
            \hline
            \multirow{2}{*}{\textbf{Parameters}} 
                & \multicolumn{2}{c}{\textbf{$\rho$ = 1}} 
                & \multicolumn{2}{c}{\textbf{$\rho$ = 0.8}} 
                & \multicolumn{2}{c}{\textbf{$\rho$ = 0.6}} 
                & \multicolumn{2}{c}{\textbf{$\rho \sim$ Unif(0.5, 0.9)}}  \\ 
            \cline{2-9}
             & \textbf{HR} & \textbf{95\% CI} 
             & \textbf{HR} & \textbf{95\% CI} 
             & \textbf{HR} & \textbf{95\% CI} 
             & \textbf{HR} & \textbf{95\% CI} \\ 
            \hline
            \multicolumn{9}{l}{\textbf{Progression-specific survival component}} \\ 
            $\log(\text{PSAD})$ & 1.48 & [1.12, 1.89] & 1.49 & [1.08, 1.94] & 1.56 & [1.15, 2.05] & 1.44 & [1.09, 1.88] \\
            $\log_2(\text{PSA} + 1)$ value & 1.21  & [0.97, 1.50] & 1.22 & [0.96, 1.53] & 1.17 & [0.91, 1.51] & 1.23 & [0.97, 1.57]\\ 
            $\log_2(\text{PSA} + 1)$ yearly change & 6.57 & [1.95, 17.11] & 6.80 & [1.95, 18.18] & 8.15 & [2.12, 23.81] & 5.36 & [1.90,15.94] \\
            \hline
            \multicolumn{9}{l}{\textbf{Treatment-specific survival component}} \\ 
            $\log(\text{PSAD})$ &  1.26 & [0.80, 1.87] & 1.27 & [0.87, 1.93] & 1.31 & [0.86, 2.07] & 1.26 & [0.83, 1.89] \\
            $\log_2(\text{PSA} + 1)$ value &  1.58 & [1.15, 2.10] & 1.54 & [1.13, 2.03] & 1.51 & [1.08, 2.07] & 1.54 & [1.14, 2.10]\\
            $\log_2(\text{PSA} + 1)$ yearly change &  12.53 & [1.10, 51.63] & 15.93 & [1.26, 70.53] & 15.48 & [1.22, 62.07] & 7.21 & [1.23, 40.68] \\ 
            \hline
            \multicolumn{9}{l}{
                \parbox{\linewidth}{
                    \vspace{1mm}
                    \small Sens: biopsy sensitivity; CI: credible interval; HR: hazard ratio; PSAD: baseline PSA density. \\
                    Note: the longitudinal components are not shown.
                }
            }
        \end{tabular}
        }
\end{table}

The estimated baseline hazards along with their credible intervals are displayed in Figure S3. Assuming a lower fixed sensitivity led to a higher baseline hazard of cancer progression with a wider credible interval. For the competing event, early treatment, this difference was more obvious between higher fixed sensitivity value (i.e., MCICJM with 100\% and 80\% sensitivity) and lower sensitivity value (i.e., MCICJMs with 60\% sensitivity), in both the estimates and credible intervals. Due to the lower degrees of freedom in the spline specification, the estimated progression- and treatment-specific baseline hazard from the MCICJM with the uniform prior were more linear than the baseline hazards in other MCICJMs with fixed sensitivities. As in Figure S3, in around year three when the most of the events were detected, the MCICJM with a prior resulted in the lowest progression- and treatment-specific baseline hazards (with lowest upper limit of their credible intervals). 

\section{Simulation}
\label{sec:simulation}

To evaluate the performance of the MCICJM recovering the true model parameters under different scenarios with regard to the (mis-)specification of the biopsy sensitivity, we performed a simulation study. The models under evaluation include 1) the correctly-specified MCICJM, 2) the MCICJM with a misspecified (fixed) sensitivity parameter, 3) a prior distribution for $\rho$ that is centred around the true sensitivity, and 4) an MCICJM with a uniform prior that covers but is not centred around the true sensitivity. The performance in estimation was assessed by comparing the coefficients between the true model (i.e., the model used to simulate the data) and the fitted models. 


\subsection{Simulation Settings}
\label{ss}
We simulated 200 datasets based on the MCICJM fitted on the Canary PASS data with a biopsy sensitivity of 75\% and the PASS biopsy schedule (see Section \ref{sec:data}). Each simulated dataset had 500 subjects. Five MCICJMs were fitted on the training data, namely an MCICJM assuming 60\% sensitivity, 75\% sensitivity and 100\% sensitivity, an MCICJM with a $\text{Unif(0.6, 0.9)}$ prior for the sensitivity parameter (i.e., centred around the true value), and an MCICJM with a sensitivity prior following $\text{Unif(0.5, 0.8)}$ (i.e., the prior still covers the true sensitivity but is not centred around it). 

We use three indicators to assess the performance of the models' estimation: the relative bias of the estimated coefficients (parameter recovery), the width of the 95\% credible interval, and the coverage of the 95\% credible intervals. 

\subsection{Simulation Results}
\label{sr}

\begin{figure}[H]
    \centering
    \subfigure[Parameter recovery (relative bias)]{\includegraphics[width=\textwidth]{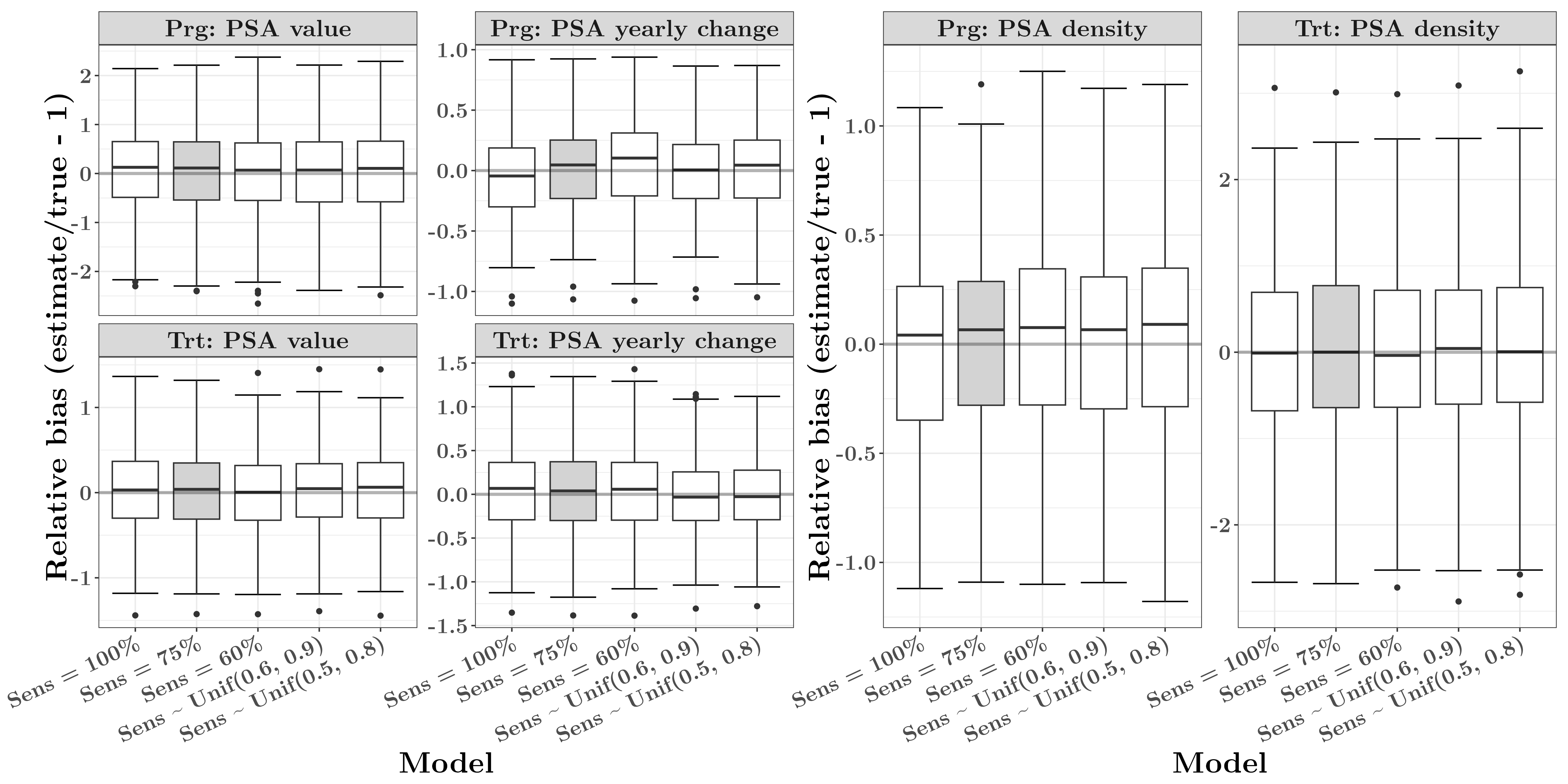}}
    \subfigure[Uncertainty (CI width)]{\includegraphics[width=\textwidth]{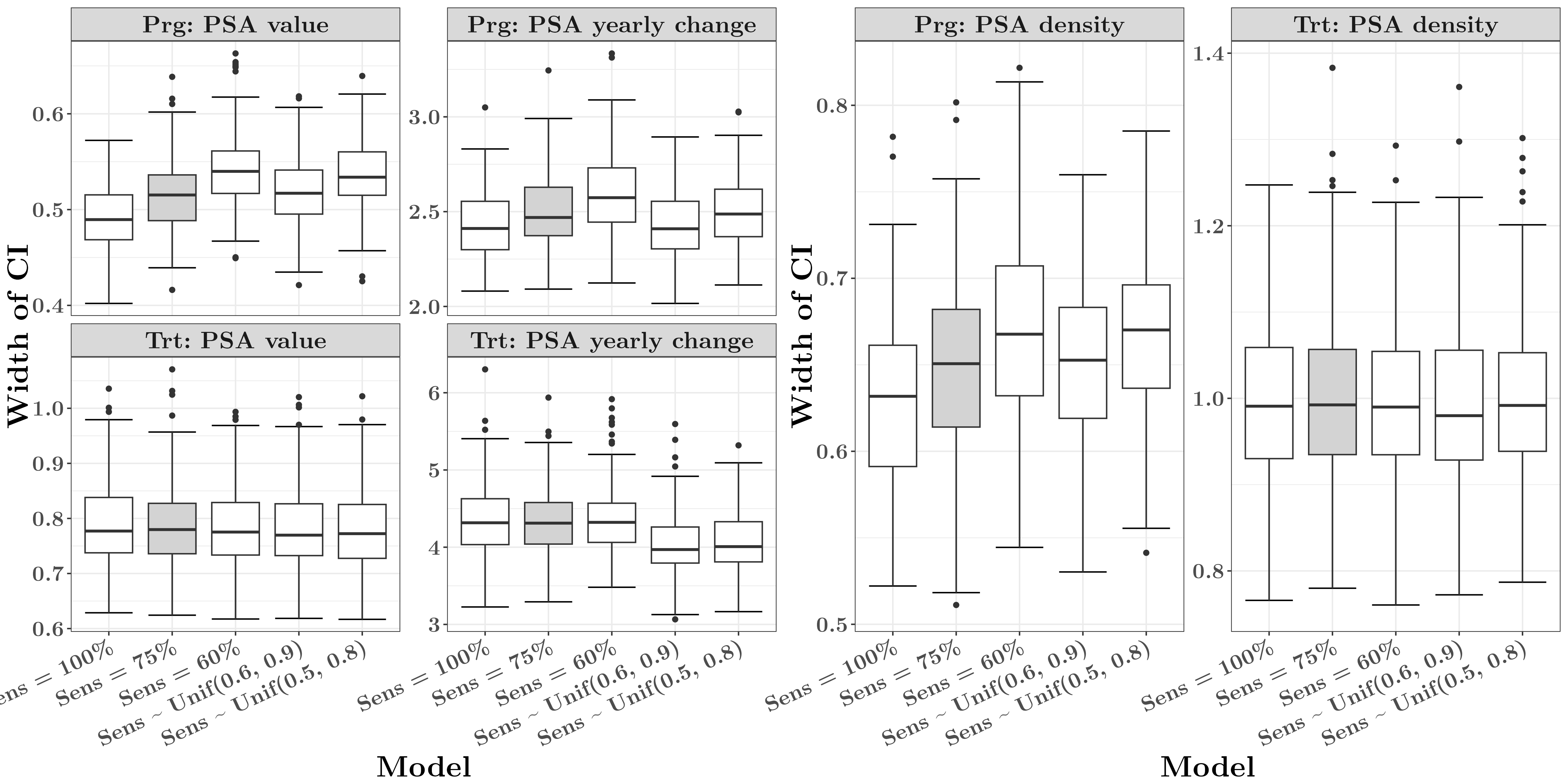}} 
   \caption{The parameter recovery (a) and uncertainty (b) of the association parameters $\boldsymbol{\alpha}$ (left) and coefficients of the baseline covariate PSA density $\boldsymbol{\gamma}$ (right) from the models with a fixed sensitivity of 60\%, 75\%, 100\%, and two models with a sensitivity prior of $\text{Unif}(0.6, 0.9)$ and $\text{Unif}(0.5, 0.8)$. The grey horizontal line indicates the true parameters. The shaded boxplot corresponds to the correctly specified model.}
   \label{fig:alphagamma}
\end{figure}

Among the $5 \times 200$ fitted models, two of the MCICJMs with 60\% sensitivity did not converge and, thus, were excluded from the analysis. The estimated average trajectories of PSA values over time from each of the five models are presented in Figure S4. All five models captured the true PSA development and its uncertainty well. The parameter recovery and credible interval width of coefficients in the time-to-event part of the MCICJMs, namely the association parameters $\boldsymbol{\alpha}$ and coefficient of baseline covariate PSA density $\boldsymbol{\gamma}$, are displayed in Figure \ref{fig:alphagamma}. The mean square errors of the above-mentioned parameters are summarized in Table~\ref{tab:msealphagamma}. All five models generated similar progression- and treatment-specific estimates and their mean square errors compared to the corresponding true parameters were resembling. The credible interval width of treatment-specific coefficients was comparable across the five models. However, the credible intervals of the progression-specific parameters were narrower in models assuming higher biopsy sensitivities than when assuming lower sensitivity. For example, the credible intervals for the MCICJM with 100\% sensitivity were, on average, 4.39\% narrower than those for the correctly-specified model (with 75\% sensitivity). While the credible intervals for the MCICJM with 60\% sensitivity were, on average, 5.22\% wider. The MCICJMs with a prior of $\text{Unif(0.6, 0.9)}$ and $\text{Unif(0.5, 0.8)}$ produced on average 0.88\% and 4.46\% wider credible intervals than the correctly-specified model. 

\begin{table}[H]
    \setlength\extrarowheight{3pt}
    \caption{The mean squared errors of the association parameters $\boldsymbol{\alpha}$ and coefficients of the baseline covariate PSA density $\boldsymbol{\gamma}$ from the models with a fixed sensitivity of 60\%, 75\%, 100\%, and two models with a sensitivity prior of $\text{Unif}(0.6, 0.9)$ and $\text{Unif}(0.5, 0.8)$.}
    \label{tab:msealphagamma}
    \resizebox{\textwidth}{!}{%
    \begin{tabular}{lccccc}
        \hline

            & \textbf{$\rho$ = 100\%} 
            & \textbf{$\rho$ = 75\%} 
            & \textbf{$\rho$ = 60\%}  
            & \textbf{$\rho \sim$ Unif(0.6, 0.9)} 
            & \textbf{$\rho \sim$ Unif(0.5, 0.8)}  \\ 
        \hline
        \multicolumn{6}{l}{\textbf{Progression-specific survival component}} \\ 
        $\log(\text{PSAD})$ & 0.031 & 0.033 & 0.036 & 0.033 & 0.035  \\
        $\log_2(\text{PSA} + 1)$ value & 0.019 & 0.021 & 0.023 & 0.021 & 0.022 \\ 
        $\log_2(\text{PSA} + 1)$ yearly change & 0.40 & 0.41 & 0.45 & 0.37 & 0.39 \\
        \hline
        \multicolumn{6}{l}{\textbf{Treatment-specific survival component}} \\ 
        $\log(\text{PSAD})$ & 0.07 & 0.069 & 0.068 & 0.07 & 0.07 \\
        $\log_2(\text{PSA} + 1)$ value & 0.039 & 0.039 & 0.040 & 0.038 & 0.038 \\
        $\log_2(\text{PSA} + 1)$ yearly change & 1.26 & 1.30 & 1.30 & 1.02 & 1.00\\ 
        \hline
        \multicolumn{6}{l}{
            \parbox{\linewidth}{
                \vspace{1mm}
                \small Sens: biopsy sensitivity; PSAD: baseline PSA density.
            }
        }
    \end{tabular}
    }
\end{table}

\begin{figure}[H]
    \centering
    \subfigure[Relative bias]{\includegraphics[width=\textwidth]{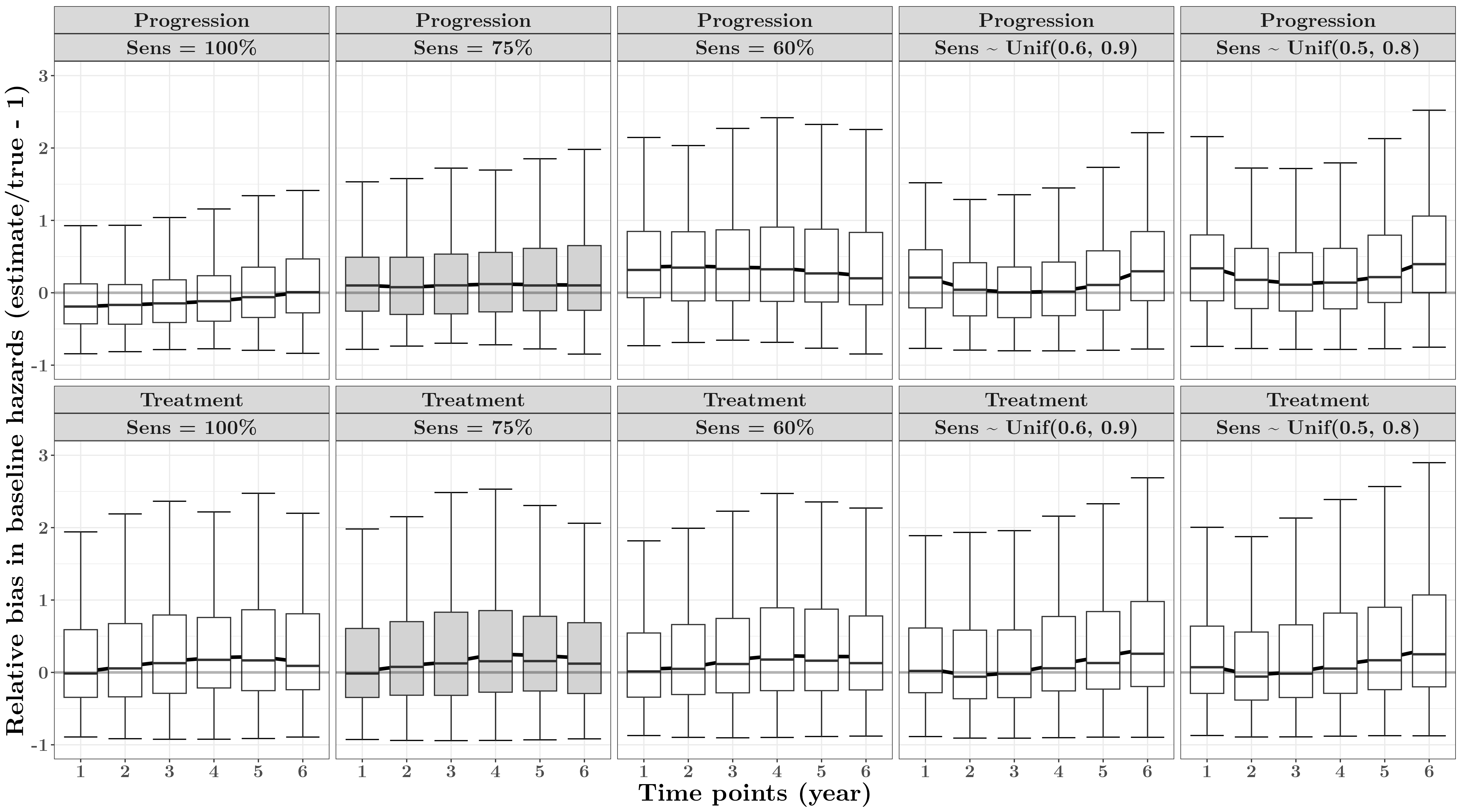}}
    \subfigure[Uncertainty (credible inerval width)]{\includegraphics[width=\textwidth]{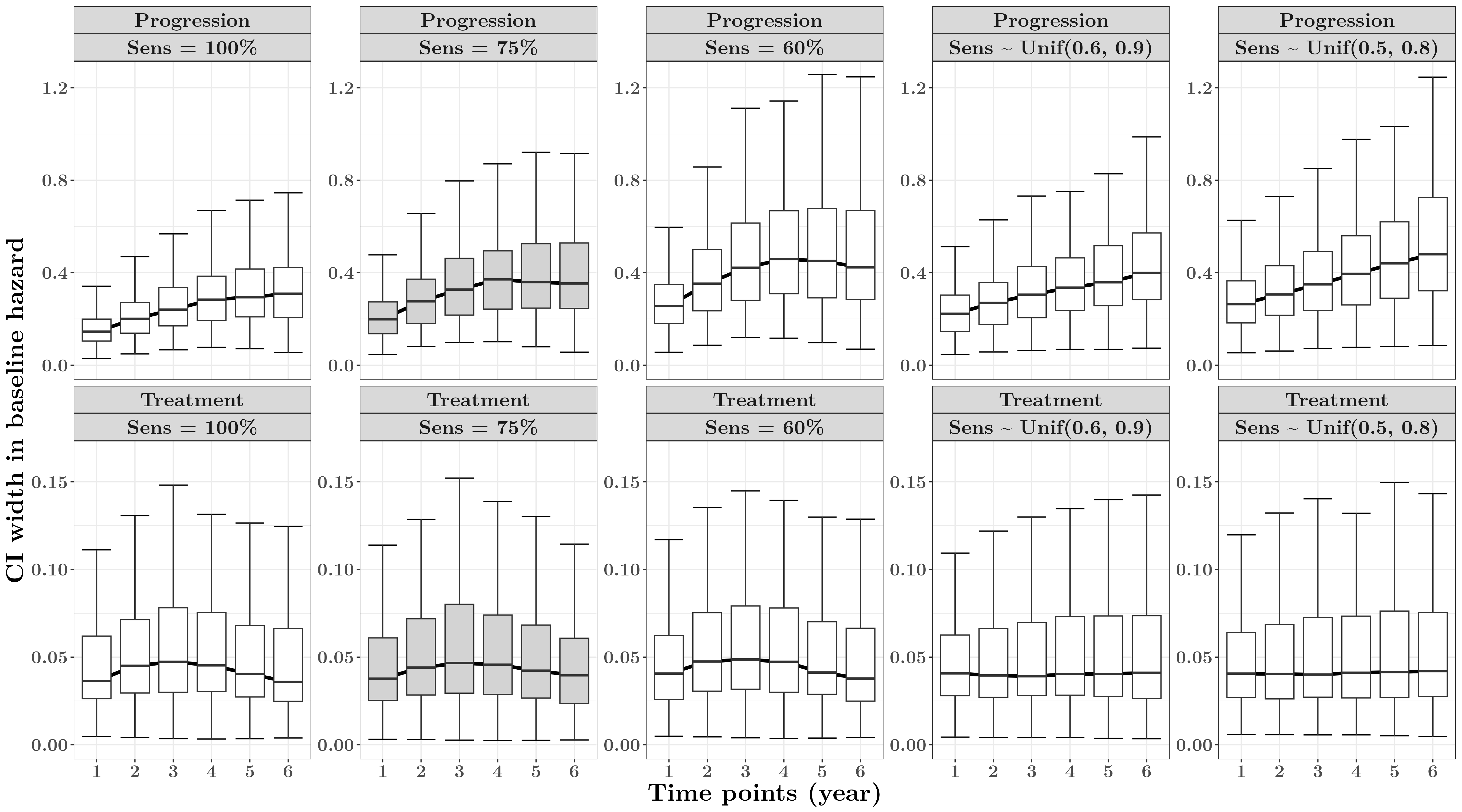}}
   \caption{The relative bias in estimates (a) and credible interval width (b) of the baseline hazards from models [MCICJMs with a fixed sensitivity of 60\%, 75\%, 100\%, and two MCICJMs with a sensitivity prior of $\text{Unif}(0.6, 0.9)$ and $\text{Unif}(0.5, 0.8)$].The shaded boxplot corresponds to the correctly specified model.}
   \label{fig:bh}
\end{figure}

The relative bias of the estimated baseline hazards and their uncertainty over time are displayed in Figure \ref{fig:bh}. A higher fixed sensitivity resulted in underestimating the progression-specific baseline hazards as well as the width of the credible interval. The estimated log-baseline hazard from the model with 100\% sensitivity was on average 14.04\% lower than the estimated hazard from the correctly specified model, whereas the MCICJM with 60\% resulted in 12.06\% higher for this hazard. In contrast, the MCICJMs with priors for the sensitivity ($\text{Unif(0.6, 0.9)}$ and $\text{Unif(0.5, 0.8)}$) performed better, leading to 2.31\% lower and 4.64\% higher estimates for the log baseline hazard, respectively. The MCICJMs with priors for biopsy sensitivity performed better in estimating the baseline hazards between year 2 and 4 where most cases occur. The baseline hazards from MCICJMs with priors for biopsy sensitivity resulted in more linear fits than those from the models with fixed sensitivities, as shown in Figure S6. 

\begin{figure}[H]
    \centering
    \includegraphics[width=\textwidth]{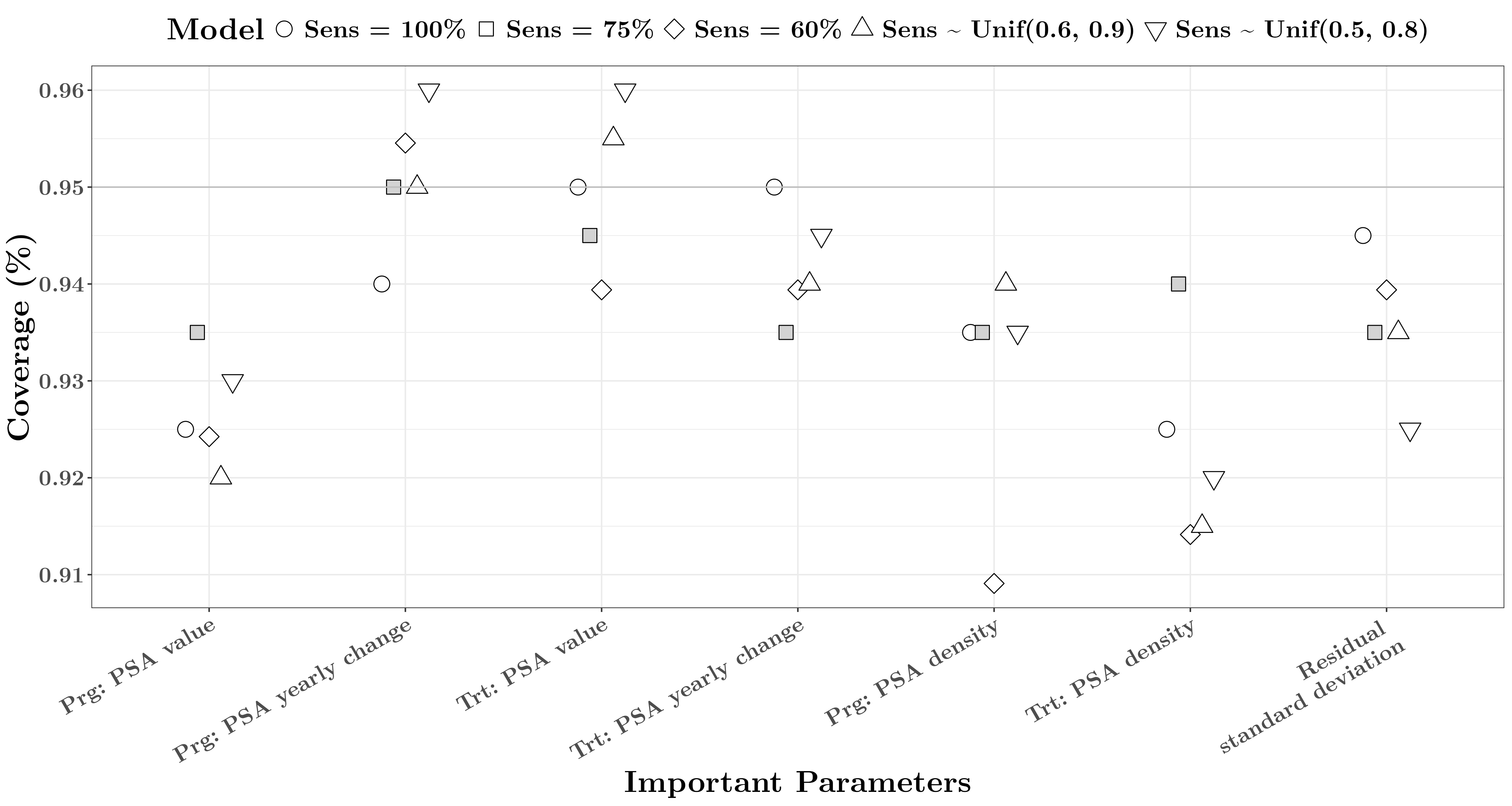}
   \caption{The percentage that the 95\% credible intervals of the important parameter estimates from the six MCICJMs fitted on the 200 simulated datasets cover the true parameters (including the $\boldsymbol{\alpha}$, $\boldsymbol{\gamma}$, and standard deviation of the residual $\epsilon$). The shaded squares correspond to the correctly specified model.}
   \label{fig:cov}
\end{figure}

The coverage of seven important parameters, namely $\boldsymbol{\alpha}$, $\boldsymbol{\gamma}$ and the standard deviation of residuals ($\epsilon$), is visualized in Figure~\ref{fig:cov}. The credible intervals of the estimated parameters in all five models covered the true parameters in more than 90\% of the replicates.

\section{Discussion}
\label{sec:discussion}

In this study, the Misclassification-Corrected Interval-censored Cause-specific Joint Model was proposed to estimate the association between the longitudinal biomarker PSA and patients' cancer progression risk, which is subject to interval censoring and misclassification, in the presence of a competing event. The hazard of the underlying event, cancer progression, is typically of more clinical relevance than the observed event, detection of progression. An extra parameter, representing the sensitivity of a biopsy to detect cancer progression, was included in the model likelihood to consider the uncertainty about the intervals during which progression happened. This parameter can either be set to a pre-specified value, or a reasonable prior can be imposed if the sensitivity is unknown. Our simulation study showed that the point estimates for the regression coefficients from our proposed model were relatively robust with regard to misspecification of the biopsy sensitivity. Nevertheless, overestimation of the biopsy sensitivity led to an underestimation of the parameter uncertainty for either the scenario when the fixed sensitivity or the prior was used. Furthermore, the assumption of sensitivities higher than the true value resulted in an underestimation of the baseline hazard, which may become an issue in the context of prediction. Using a uniform prior for the biopsy sensitivity performed well in estimating the baseline hazard in our case and may, thus, mitigate the bias in prediction caused by a misspecified biopsy sensitivity. 

The application of the MCICJM is not only limited to our specific case of prostate cancer. The issues of interval censoring due to periodic examinations and the potential consequence of misclassification are prevalent in the medical field, such as screening tests in oncology and infectious diseases. Moreover, in many applications, death is a competing event to the outcomes of interest.

In applying the proposed methodology, there are two restrictions to be noted. Firstly, the use of a prior distribution for the biopsy sensitivity in the MCICJM, to handle the situation when the exact sensitivity is unknown, gives rise to identifiability issues. This requires restrictions to be imposed on the flexibility of the baseline hazard. Furthermore, in this paper, we assumed that biopsies cannot be false-positive, and that the biopsies performed at baseline to identify patients eligible for active surveillance are error-free, i.e., none of the included patients had progressed cancer at the start of active surveillance. In the Canary PASS, AS patients underwent two biopsies to confirm the absence of cancer progression before they entered the cohort, which reduces the risk of misclassification at baseline. Nevertheless, in other applications this might not be the case, leading to left truncation.\citep{Applebaum2011} In our model, this could be considered by introducing an offset previously mentioned in Section~\ref{subsec:notation}. To determine this offset, we need to specify a risk of progression before AS. Since there is no valuable information prior to the baseline, this term cannot be explicitly modeled. Furthermore, in our case, this value is rarely disclosed in the previous literature.

Further research will focus on the clinical applications of the proposed MCICJM. For example, the model can potentially be used as a dynamic prediction tool to forecast the risk of patients experiencing underlying biological cancer progression that can never be observed in practice. In the earlier work, we used  joint modelling (without considering biopsy sensitivities) to provide patient-specific AS schedules based on this predicted risk.\citep{Tomer2022, Yang2023} By considering the imperfection of the examinations, the model is expected to offer an improvement over the existing methodology in such real-world applications. 

In conclusion, working in the Bayesian framework allowed us to further extend the joint modelling methodology to take into account the uncertainty brought by imperfect examinations. This uncertainty can either be flexibly modelled by a fixed sensitivity constant at the risk of misspecification, or considered by a sensitivity prior to reduce the risk of misspecification but sacrificing the model flexibility. The trade-off between them needs to be decided.

\section*{Acknowledgements}
The research was funded by the National Institutes of Health (the NIH CISNET Prostate Award CA253910). The authors would also like to show our gratitude to the Canary PASS team and all study participants.

\section*{Conflict of interest}
The authors declare no potential conflict of interests.

\bibliographystyle{abbrvnat}
\bibliography{myref}




\end{document}


\maketitle

\section{Data}

\subsection{PASS Data}

Table~\ref{Tab:t1} summarizes the relevant subset of the Canary PASS data. 

\begin{table*}[!ht]
    \centering
    \caption{Summary table for the Canary PASS Data.}
    \begin{tabular}{ll}
        \toprule
        \textbf{Item}                            & \textbf{Value} \\ \midrule
        Number of subjects  & 833  \\
        Observation time until progression/treatment (years)$^*$    & 4.35 (2.82-6.18)  \\
        Baseline PSA density$^\ddagger$ ($\text{ng}/\text{ml}^2$)$^\dagger$  & 0.12 (0.10)  \\
        Age at start of AS (years)$^*$  & 62 (57-67)   \\
        Total number of PSA measurements & 8262  \\
        Number of PSA measurements per patient$^*$   & 9 (5-14) \\
        PSA level (ng/ml)$^\dagger$  & 5.10 (3.84)  \\
        Number of positive cores per patient$^*$ & 3 (2-4) \\
        core ratio (\%)$^*$ & 8.33 (0.00-16.67)\\
        Number of biopsies per patient$^*$ & 2 (2-3)  \\
        \bottomrule
        \multicolumn{2}{l}{\small $^*$ median is shown followed by the interval between 25\% quantile and 75\% quantile; } \\
        \multicolumn{2}{l}{\small $^\dagger$ mean is shown with standard deviation in the brackets; } \\
        \multicolumn{2}{l}{\small $^\ddagger$: PSA density equals to PSA level (ng/ml) divided by prostate volume (ml).}
    \end{tabular}
    \label{Tab:t1}
\end{table*}

\subsection{Time-to-event Outcomes}

\begin{figure}[H]
    \centering
    \includegraphics[width = 0.9\textwidth]{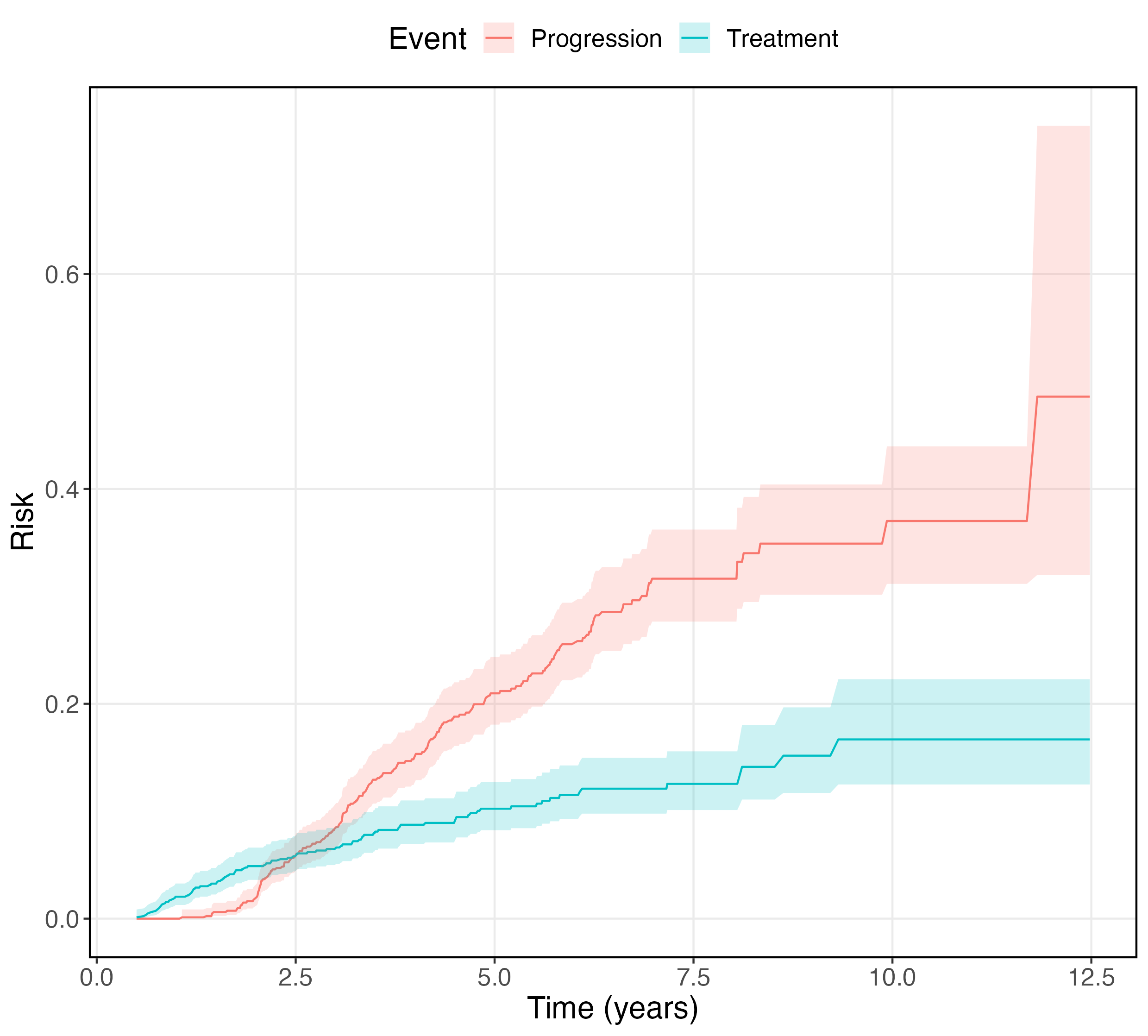}
    \caption{The Aalen-Johansen estimator of progression- and treatment-specific risk in the PASS data.}
    \label{fig:aj}
\end{figure}

The Aalen–Johansen estrimator of the two events' risks in the PASS data is visualized in Fig.~\ref{fig:aj}.

\subsection{Longitudinal Outcomes}

\begin{figure}[H]
    \centering
    \includegraphics[width = 0.9\textwidth]{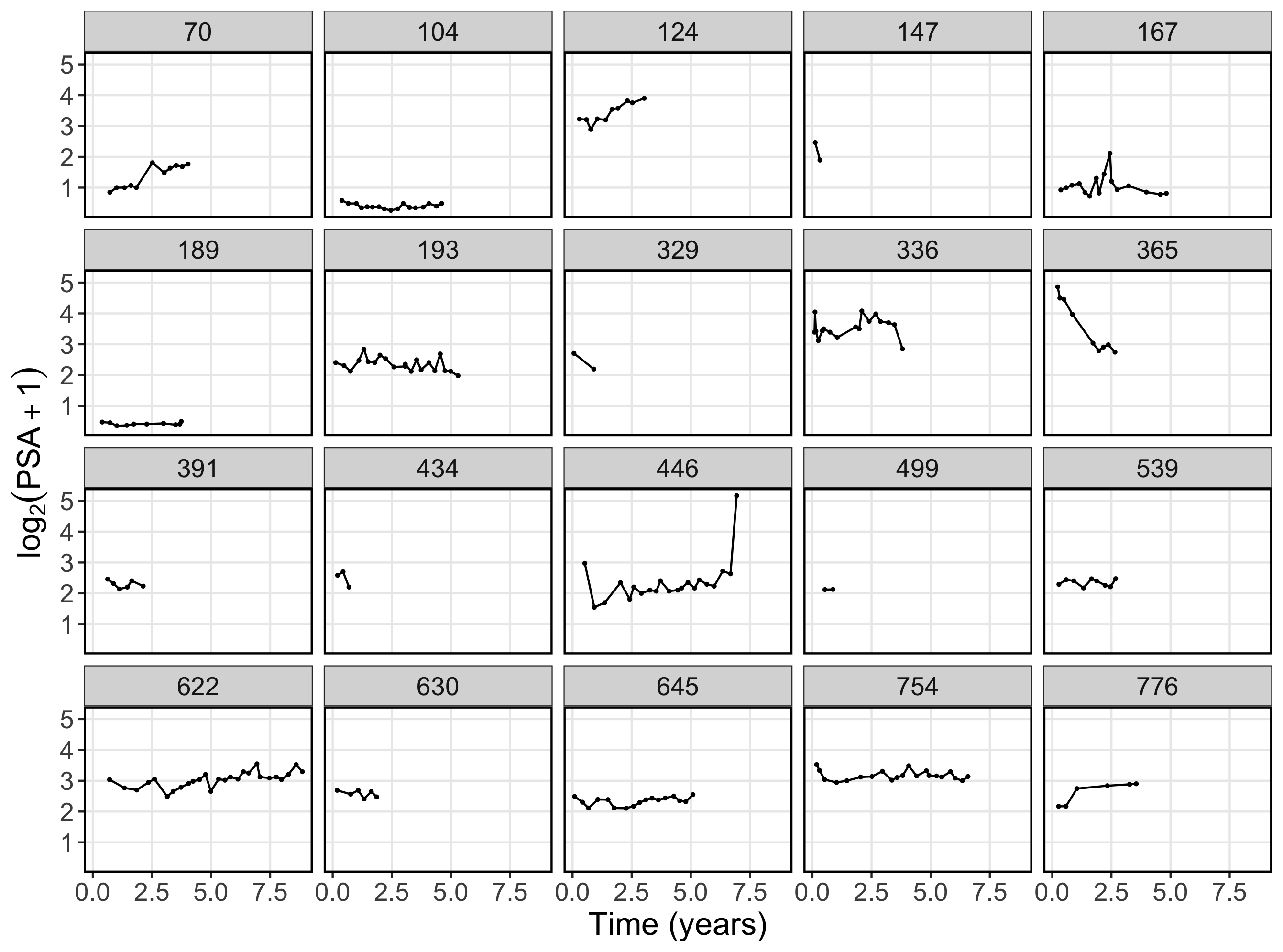}
    \caption{Observed trajectories of PSA for 20 randomly selected subjects.}
    \label{fig:observedtrajectory}
\end{figure}

In the Canary PASS data, two longitudinal outcome are available, namely, PSA levels and proportion of cores obtained by a biopsy that contain cancerous cells (core ratio). In Fig.~\ref{fig:observedtrajectory}, the development of the longitudinal outcomes is displayed for 20 randomly selected patients. The trajectories show non-linear evolutions over time, and vary greatly between patients, which needs to be accommodated in the longitudinal component of the ICJM.

\section{Misclassification-Corrected Interval-censored Cause-specific Joint Models (MCICJM)}

In the specification of the MCICJM, we use a flexible semi-parametric specification of the baseline hazard using penalized B-splines,
\begin{align*}
    \log h_0^{(k)}(t) = \gamma_{k, h_0, 0} + \sum^A_{a=1}{\gamma_{k, h_0, a}\mathcal{G}_a(t, \boldsymbol{\xi})},
\end{align*}
where $\mathcal{G}_a(t, \boldsymbol{\xi})$ is the $a$-th basis function of a B-splines with knots $\xi_1, \dots, \xi_A$. The number of knots was chosen to be 11 for MCICJMs with fixed sensitivity $\rho$ and 4 for MCICJMs with a sensitivity prior. The penalized coefficients for the basis function $\boldsymbol{\gamma}_{k, h_0}$ have the following priors,
\begin{align*}
    p(\boldsymbol{\gamma}_{k, h_0} \mid \tau_{k,h_0}) \propto \tau_{k,h_0}^{\rho(\boldsymbol{M})/2} \exp\left(- \frac{\tau_{k,h_0}}{2}\boldsymbol{\gamma}_{k, h_0}^\top\boldsymbol{M}\boldsymbol{\gamma}_{k, h_0}\right),
\end{align*}
with
\begin{align*}
     \tau_{k,h_0} \sim \text{Gamma}(5, 0.5),
\end{align*}
where $\tau_{k,h_0}$ is the smoothing parameter; $\boldsymbol{M} = \Delta_r^\top \Delta_r + 10^{-6}I$, $\Delta_r$ is the $r$-th difference penalty matrix and $\rho(\boldsymbol{M})$ denotes the rank of $\boldsymbol{M}$. 

\section{MCICJMs for Real Data Analysis}

\begin{figure}[H]
   \includegraphics[width=0.9\textwidth]{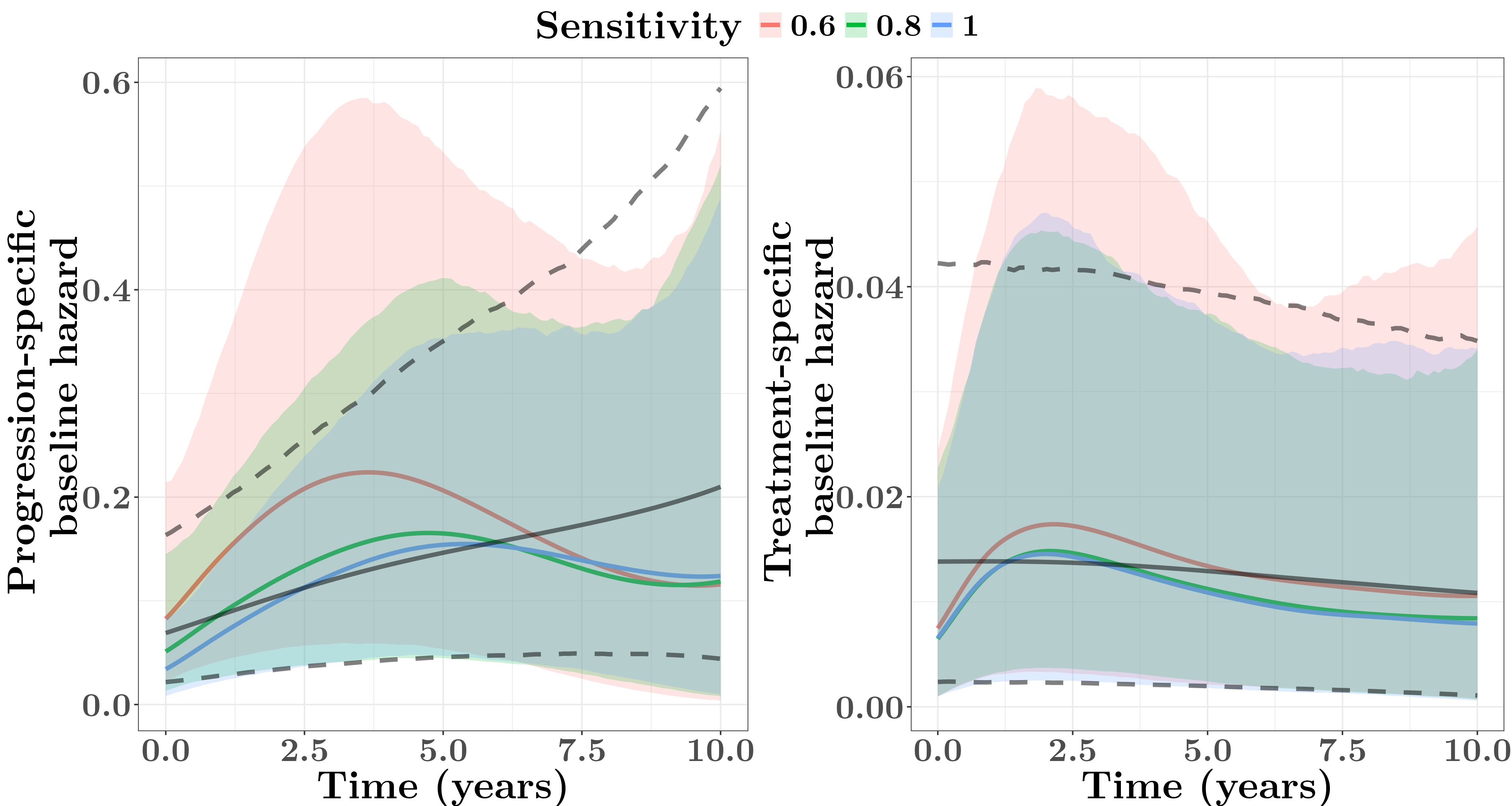}
   \caption{Estimated progression-specific (left) and treatment-specific (right) baseline hazard from the three MCICJMs with different fixed sensitivity values (in color) and the MCICJM with a uniform sensitivity prior between 0.5 and 0.9 (in black; the solid and dashed lines indicate the posterior mean and 2.5\% and 97.5\% quantiles, respectively).}
   \label{fig:res_bh}
\end{figure}

The baseline hazards and the corresponding 95\% credible intervals of the four MCICJMs fitted on the PASS data are presented in Fig.~\ref{fig:res_bh}.

\section{Simulation Study}

\subsection{Simulation Setting} \label{subsection:icjm1}

To evaluate the performance of our proposed methodology, we simulated data based on the MCICJM with a fixed biopsy sensitivity of 75\% fitted on the Canary PASS study. The patients in the training sets are supposed to take biopsies at months 12, 24 and afterwards biennially and PSA measurements every three months, with small variations.

This model had the following structure:
\begin{align*}
    \log_2\{\text{PSA}_i(t) + 1\} &= m_{i}(t) + \epsilon_i(t),\\
                    m_{i}(t)  &= \beta_0 + u_{0i} + \sum^3_{p=1}(\beta_p + u_{pi})\mathcal{C}^{(p)}_i(t) + \beta_4(\text{Age}_i - 62), \\
    h_i^{(k)}\left\{t \mid \boldsymbol{\mathcal{M}}_{i}(t)\right\} &= h_0^{(k)}(t)\exp\Big[\gamma_k\text{density}_i + f\left\{\boldsymbol{\mathcal{M}}_{i}(t), \boldsymbol{\alpha}_{k}\right\}\Big],
\end{align*}
where $\mathcal{C}(t)$is the design matrix for the natural cubic splines (with three degrees of freedom) for time $t$; $\text{Age}_i$ and $\text{density}_i$ refer to the patient's age and PSA density at the start of active surveillance, respectively. Baseline Age was centred by subtracting the median age (62 years) for computational reasons. Both the expected value of PSA and the change in expected PSA over the previous year (where extrapolation was conducted for time points earlier than year one) were included as covariates in the time-to-event component, i.e.,
\begin{align*}
    f\left\{\boldsymbol{\mathcal{M}}_{i}(t), \boldsymbol{\alpha}_{k}\right\} = &\  \alpha_{1k}m_{\textsc{psa},i}(t) + \alpha_{2k}\Big\{m_{\textsc{psa},i}(t) - m_{i}(t-1)\Big\}.
\end{align*}

The residuals of the longitudinal component were assumed to follow a student's t distribution with three degrees of freedom \cite{Tomer2022},
\begin{align*}
    \epsilon_i(t) \sim t\left(\frac{1}{\tau_\epsilon}, 3\right),
\end{align*}
with
\begin{align*}
    \tau_\epsilon \sim \text{Gamma}(0.01, 0.01).
\end{align*}
The prior distributions for the regression coefficients were specified as vague normal distributions,
\begin{align*}
    \beta &\sim \mathcal{N}(0, 100), \\
    \gamma_k &\sim \mathcal{N}(0, 100), \\
    \alpha_{1k}, \alpha_{2k} &\sim \mathcal{N}(0, 100),
\end{align*}
and the variance-covariance matrix of the random effects, $\boldsymbol{\Omega}$, to follow an inverse-Wishart distribution,
\begin{align*}
    \boldsymbol{\Omega} \sim \mathcal{IW}(n_u + 1, \frac{4}{\tau_u}),
\end{align*}
with
\begin{align*}
    \tau_u \sim \text{Gamma}(0.5, 0.01),
\end{align*}
where $n_u$ is the number of random effects per subject.

The model was implemented in JAGS \cite{JAGS} and run for 10000 iterations, using a thinning interval of 10, in each of three MCMC chains.

The resulting posterior means used for simulation were
\begin{align*}
    \boldsymbol{\beta} &= [2.35, 0.27, 0.62, 1.00, 0.02]^\top, \\
    \boldsymbol{\Omega} &= \begin{bmatrix}
        0.49 & -0.04 & -0.09 & 0.00 \\
        -0.04 & 0.77 & 0.43 & -0.08 \\
        -0.09 & 0.43 & 1.41 & 1.43 \\
        0.02 & -0.08 & 1.43 & 2.60
    \end{bmatrix}, \\
    \tau_\epsilon &= 47.39, \\
    \boldsymbol{\gamma}_{h_0} &= \begin{bmatrix}
        -3.02 & -5.13 \\
        -2.57 & -4.55 \\
        -2.17 & -4.26 \\
        -1.87 & -4.31 \\
        -1.78 & -4.47 \\
        -1.87 & -4.65 \\
        -2.04 & -4.80 \\
        -2.26 & -4.91 \\
        -2.51 & -5.11 \\
        -2.74 & -5.37 \\
        -2.95 & -5.66 \\
        -3.15 & -5.97
    \end{bmatrix},\\
    \boldsymbol{\gamma} &= [0.41, 0.25], \\
    \boldsymbol{\alpha} &= \begin{bmatrix} 
        0.16 & 0.40 \\
        1.79 & 2.22
    \end{bmatrix}.
\end{align*}

The resulting simulated data matched the observed data well with regard to the rates of cancer progression, early treatment initiation and censoring (Table~\ref{Tab:summarytraining}).

\begin{table}[H]
    \centering
    \caption{Summary of event proportions in the simulated training datasets compared to the observed data.}
    \begin{tabular}{lcc}
        \toprule
        \textbf{Events}  & \textbf{Simulated data}$^\dagger$ (\%) & \textbf{Observed data} (\%) \\ \midrule
        Cancer progression & 22.34 & 21.97 \\
        Treatment & 8.80 & 10.44 \\
        Censoring & 68.86 & 67.59 \\
        \bottomrule
        \multicolumn{3}{l}{$^\dagger$: the average proportions over all training sets are presented.}
    \end{tabular}
    \label{Tab:summarytraining}
\end{table}

\subsection{Simulation Results - Longitudinal Part}

\begin{figure}[H]
    \centering
    \includegraphics[width = 0.9\textwidth]{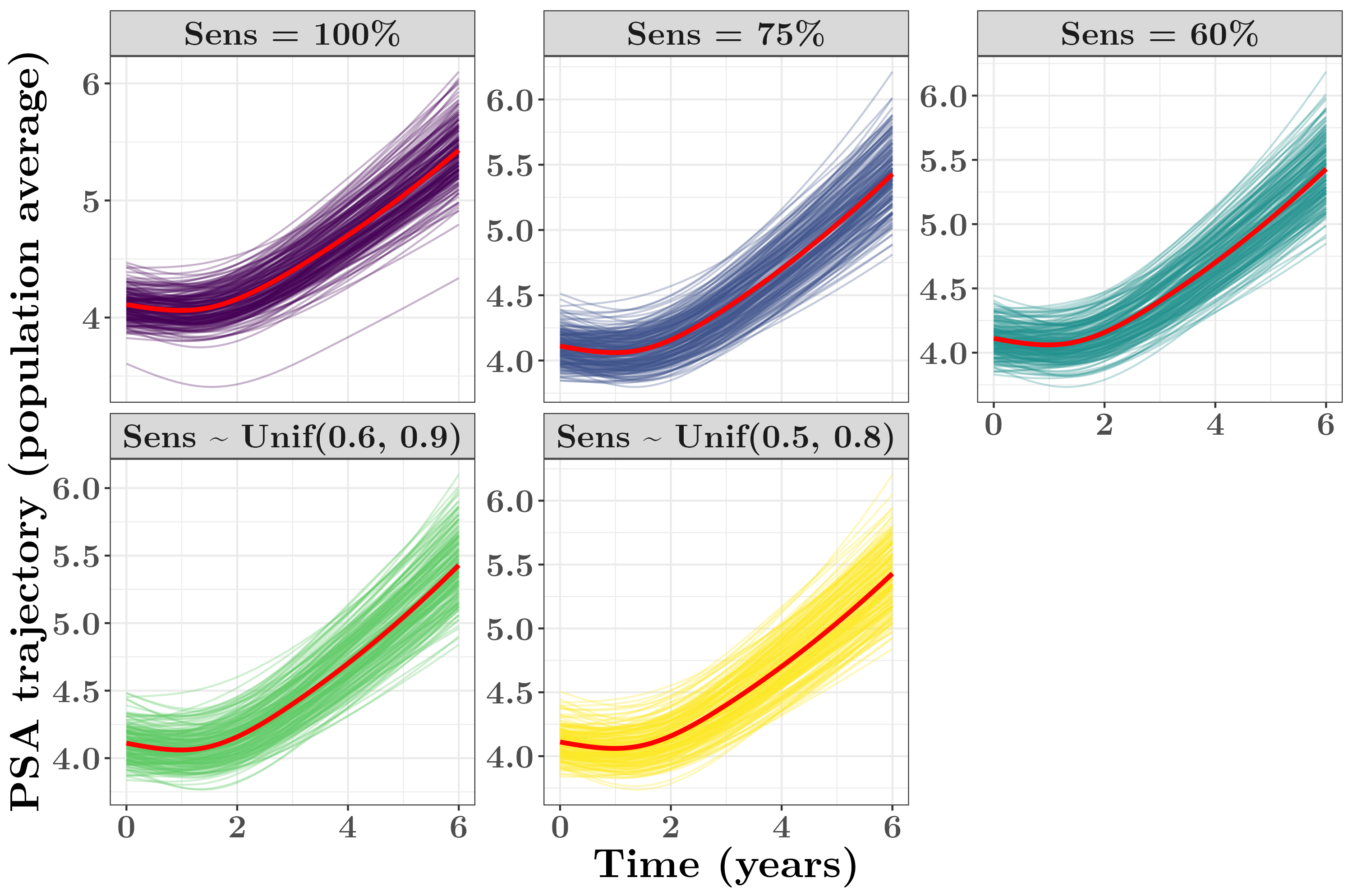}
    \caption{The estimates of the population average PSA trajectories from three MCICJMs with a fixed sensitivity of 60\%, 75\%, 100\%, and two MCICJMs with a sensitivity prior of $\text{Unif}(0.6, 0.9)$ and $\text{Unif}(0.5, 0.8)$, the thick curve indicating the true underlying trajectory.}
    \label{fig:psa_estimate}
\end{figure}

\begin{figure}[H]
    \centering
    \includegraphics[width = 0.9\textwidth]{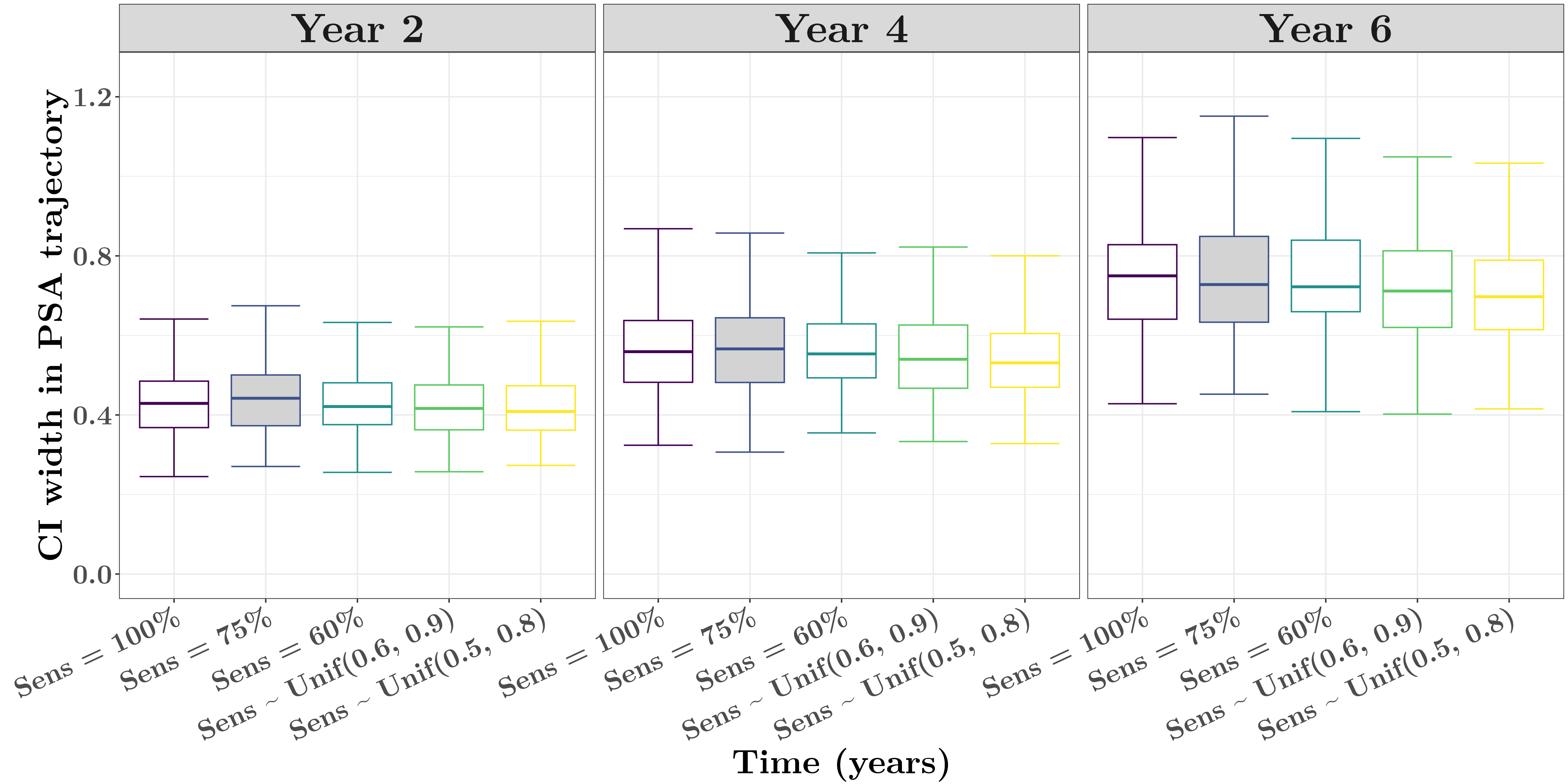}
    \caption{The credible width of the population average PSA trajectories at year 2, 4 and 6 from three MCICJMs with a fixed sensitivity of 60\%, 75\%, 100\%, and two MCICJMs with a sensitivity prior of $\text{Unif}(0.6, 0.9)$ and $\text{Unif}(0.5, 0.8)$. The results from the correctly specified model are indicated with gray shading.}
    \label{fig:psa_unc}
\end{figure}

The estimated population average trajectories of PSA values for all five models are displayed in Fig.~\ref{fig:psa_estimate}. The corresponding uncertainty, the width of the 95\% credible intervals at year 2, 4, and 6 are presented in Fig.~\ref{fig:psa_unc}. 

\subsection{Simulation Results - baseline hazards}

\begin{figure}[H]
    \centering
    \includegraphics[width = \textwidth]{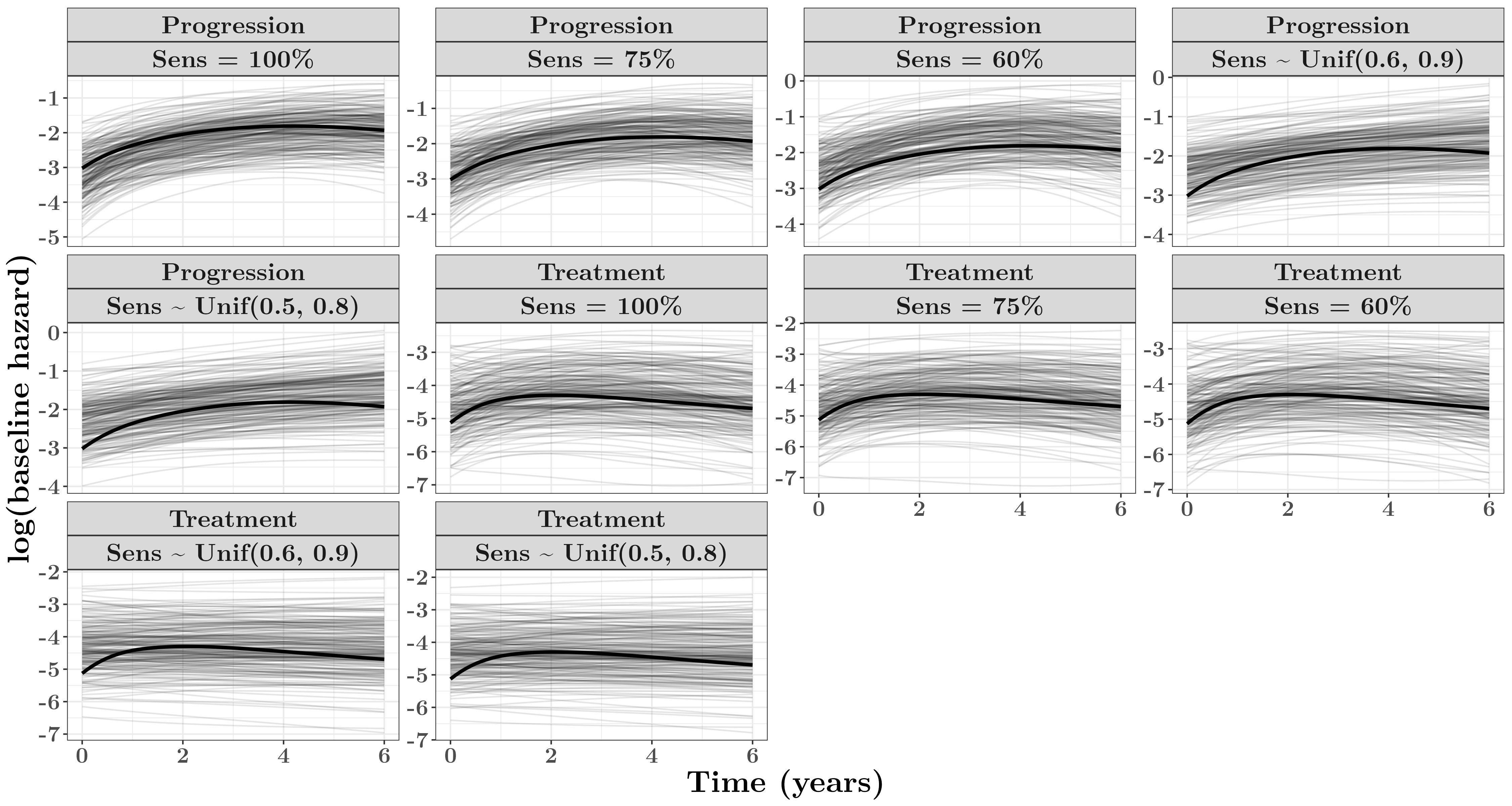}
    \caption{The estimates of the baseline hazards from models [MCICJMs with a fixed sensitivity of 60\%, 75\%, 100\%, and two MCICJMs with a sensitivity prior of Unif(0.6,0.9) and Unif(0.5,0.8)], the thick curve indicating the true baseline hazards.}
    \label{fig:bh_curv}
\end{figure}

The estimated baseline hazards in log transformation are visualized in Fig.~\ref{fig:bh_curv}.

\bibliographystyle{plain}
\bibliography{myref}